\documentstyle[twoside, epsfig]{article}

\input ibvs2.sty

\begin{document}

\IBVShead{5982}{25 March 2011}

\IBVStitletl{The Eclipsing Cataclysmic Variables PHL 1445}{and GALEX J003535.7+462353}

\IBVSauth{Wils, Patrick$^1$; Krajci, Tom$^2$; Hambsch, Franz-Josef$^{1,3}$; Muyllaert, Eddy$^1$}
\IBVSinst{Vereniging Voor Sterrenkunde, Belgium; e-mail: patrickwils@yahoo.com}
\IBVSinst{Astrokolkhoz, New Mexico, USA}
\IBVSinst{Bundesdeutsche Arbeitsge\-mein\-schaft f\"{u}r Ver\"{a}nderliche Sterne e.V. Germany} 

\SIMBADobj{PHL 1445}
\SIMBADobj{GALEX J003535.7+462353}

\IBVStyp{CV, UG, E}
\IBVSkey{Variable stars, Cataclysmic variables, Dwarf novae, Eclipsing binaries}
\IBVSabs{PHL 1445 is found to be an eclipsing cataclysmic variable with an orbital period of 76.3 minutes.  GALEX J003535.7+462353 is a new eclipsing dwarf nova with an orbital period of 4.13 hours.  Both objects show deep eclipses with an amplitude of more than two magnitudes.}

\begintext

Eclipsing cataclysmic variables (CVs) are important because through detailed modeling of the eclipses 
it is possible to deduce the physical properties of the system.  
This paper reports the discovery of two new eclipsing CVs: PHL\,1445 and GALEX\,J003535.7+462353.

PHL\,1445 (= PB\,9151) is listed in the Palomar-Haro-Luyten catalogue as a faint blue object (Haro \& Luyten, 1962).
A spectrum (6dFGS\,g0242429-114646) taken by the 6dF Galaxy Survey (Jones et al., 2004 and 2009) showed it to be a cataclysmic variable (Wils, 2009).
Because of the split emission lines and a number of anomalously faint points in the light curve of the 
Catalina Real-time Transient Survey (CRTS; Drake et al., 2009), it was suspected to be an eclipsing variable as well.
Follow-up observations at the Astrokolkhoz Observatory with a C14 Schmidt-Cassegrain and an unfiltered CCD camera, showed this indeed to be the case.  
As shown in Fig.~1, the light curve shows deep eclipses lasting about 6 minutes, with an amplitude of more than two magnitudes.
In addition the period is very short, 76.3 minutes, near the minimum orbital period for CVs (G\"ansicke et al., 2009).
Such a short orbital period is usually observed in WZ\,Sagittae type dwarf novae like GW\,Lib (orbital period 76.8 minutes) and SDSS\,J074531.91+453829.5 (76.0 minutes),
with rare large amplitude outbursts.  Only SDSS\,J150722.30+523039.8 has a shorter orbital period among the eclipsing CVs (Savoury et al., 2011).

Table~\ref{ToM} lists the observed times of eclipses.
From these, the following eclipse ephemeris was derived:
\begin{equation}
\label{PHL1445}
  HJD Min = 2455202.5579(1) + 0\fday05298466(8) \times E
\end{equation}

Since not many deeply eclipsing CVs are known at this orbital period, high speed photometry of the eclipses, 
such as done by Southworth and Copperwheat (2011) and Savoury et al. (2011) would certainly be of value for this object.

\IBVSfig{9.5cm}{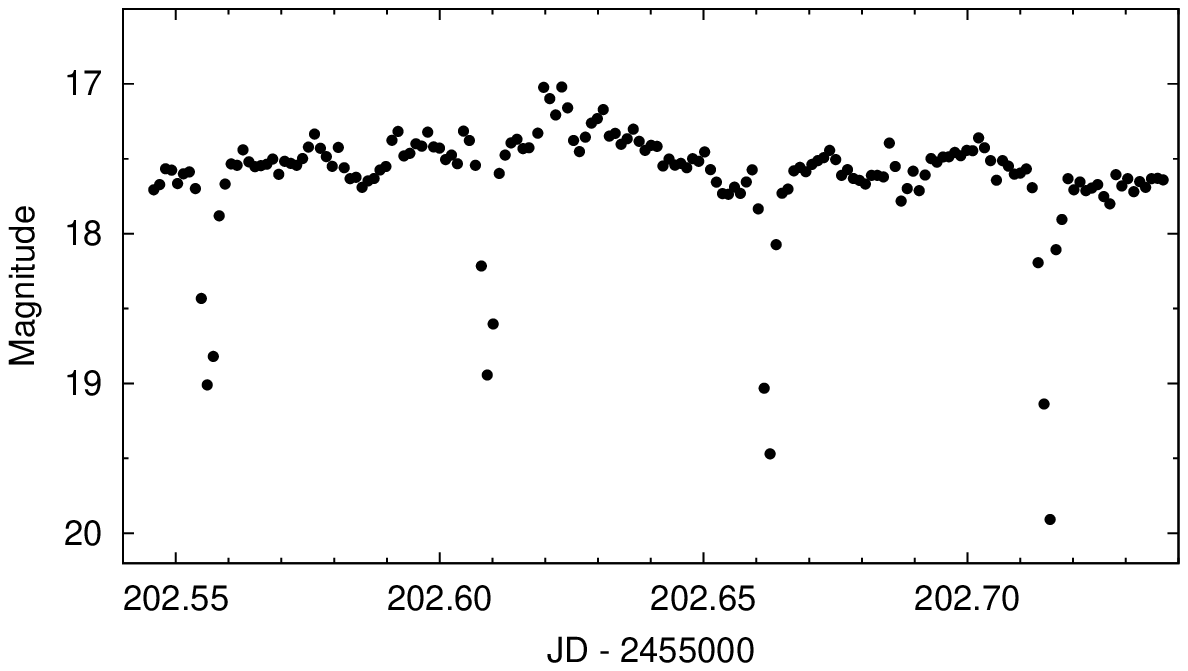}{Light curve of PHL\,1445 showing four eclipses.}
\IBVSfigKey{Ecl-CV-Fig1.eps}{PHL\,1445}{Light curve}

\begin{table}
\begin{center}
\caption{Observed times of eclipse for PHL\,1445 and GALEX\,J003535.7+462353.  
The times are given as HJD - 2450000 (UTC based).
The uncertainty on the times is about 0.0001 days for PHL\,1445 and 0.0005 days for GALEX\,J003535.7+462353 for the minima obtained from our data,
and 0.001 days for the minima obtained from SuperWASP data.} \vskip2mm
\label{ToM}
\begin{tabular}{c|c|c} 
\hline
PHL\,1445 & \multicolumn{2}{c}{GALEX\,J003535.7+462353} \\
          & SuperWASP & This paper \\
\hline
5202.5579 & 4330.553 & 5477.5621 \\
5202.6108 & 4331.589 & 5478.4228 \\
5202.6640 & 4332.622 & 5478.5954 \\
5202.7169 & 4333.655 & 5479.4560 \\
5241.6075 & 4334.688 & 5479.6284 \\
5242.6144 & 4335.551 & 5480.6625 \\
& 4360.703 & 5481.3519 \\
& 4407.388 & 5481.5239 \\
& 4408.424 & 5482.3856 \\
& & 5483.4192 \\
& & 5486.6920 \\
& & 5495.3052 \\
& & 5495.6516 \\
& & 5576.6190 \\
& & 5577.6526 \\
& & 5579.7207 \\
\hline
\end{tabular}
\end{center}
\end{table}

GALEX\,J003535.7+462353 was discovered as a variable source by the GALEX satellite (Martin et al., 2005) on 30 August 2008. \IBVSshortlink{http://www.galex.caltech.edu/researcher/tdsalerts.html}{GALEX TDA}
Although the object is too faint itself, both the Northern Sky Variability Survey (NSVS; Wo\'zniak et al., 2004) and SuperWASP (Butters et al., 2010) 
observed the combined magnitude of GALEX\,J003535.7+462353 and GSC\,3249-1603, which lies some $18\arcs$ to the West.
Both surveys show a number of brightenings in the combined light curve, lasting several days, 
with an amplitude of up to 0.2 magnitudes from the normal combined magnitude of 12.9, 
indicating the possible variability of GALEX\,J003535.7+462353 rising to about magnitude 14.5, from its normal magnitude of around 16.5.
These may be an indication of a dwarf nova outburst with a fairly small amplitude.
In addition, during these bright phases SuperWASP showed short periodic dimmings back to the normal combined magnitude 
with a period of around 0.1723 days.  
The likely cause of these periodic fadings are eclipses of the variable.

GALEX\,J003535.7+462353 was therefore followed extensively by the authors.  
The eclipses with a duration of about 30 minutes, could be easily confirmed.  
At quiescence the eclipse depth is about 2 magnitudes in $V$, but varying slightly. 
In a timespan of three months one definite outburst was observed, lasting about a week (see Fig.~2), and possibly a few shorter outbursts.  
At the end of the observing season, the object was entering another outburst.
The rise to outburst seems to be more gradual, like in some other dwarf novae with a short outburst cycle and relatively small amplitude
(often classified as Z Cam type variables).
During the long outburst, the eclipses could also be observed with a similar amplitude as during quiescence.  
Fig.~3 shows eclipses observed during quiescence, during a rise to outburst and one during outburst.

From the list of observed times of eclipse in Table~\ref{ToM},
together with the times of minimum that could be derived from the SuperWASP data, 
the following eclipse ephemeris was deduced:

\begin{equation}
\label{J003535}
  HJD Min = 2455477.5615(4) + 0\fday17227503(11) \times E
\end{equation}

\IBVSfig{9.5cm}{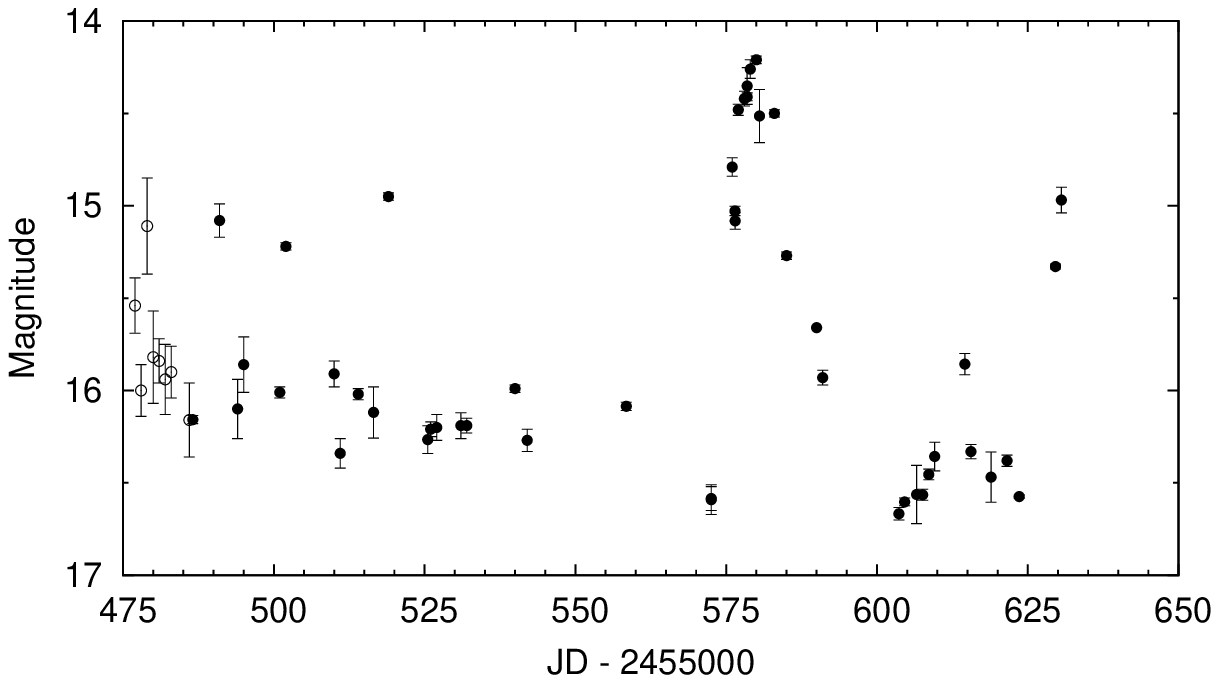}{Light curve of GALEX\,J003535.7+462353 composed of daily means of observations outside of eclipse.  
Open circles represent $V$ magnitudes, filled circles unfiltered magnitudes.}
\IBVSfigKey{Ecl-CV-Fig2.eps}{GALEX\,J003535.7+462353}{Light curve}

\IBVSfig{7cm}{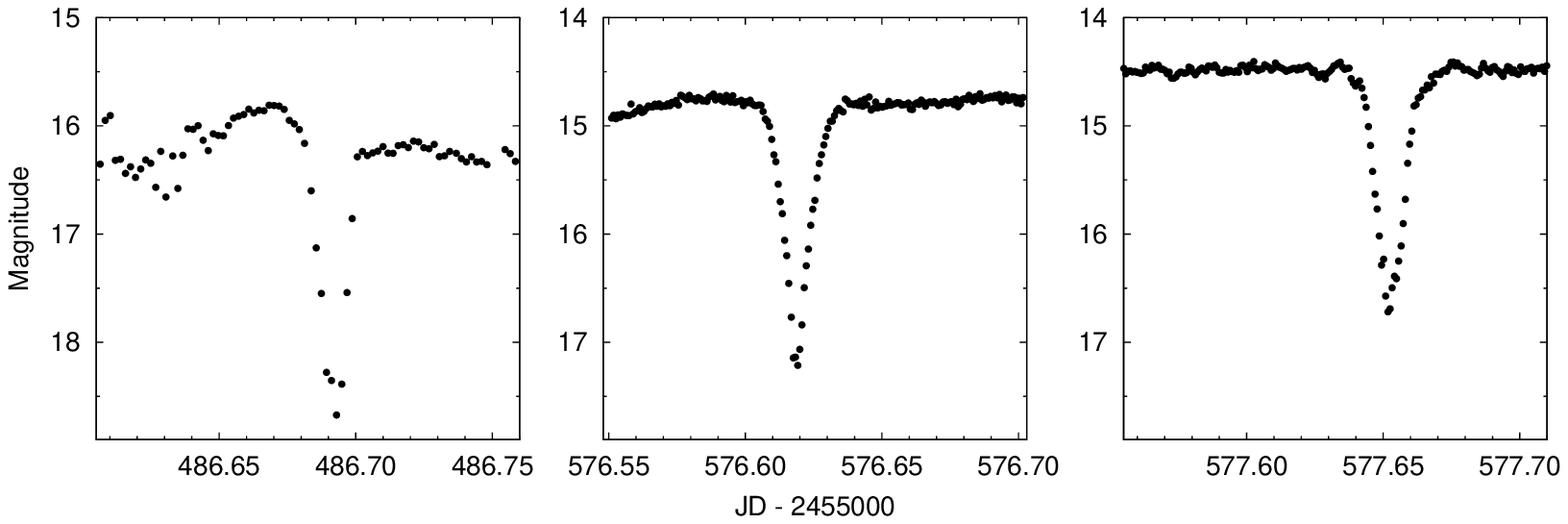}{Eclipses of GALEX\,J003535.7+462353 observed in quiescence (left), rising to outburst (middle) and in outburst (right).}
\IBVSfigKey{Ecl-CV-Fig3.eps}{GALEX\,J003535.7+462353}{Light curve}

{\bf Acknowledgements:}
This study made use of the Simbad and VizieR databases (Ochsenbein et al., 2000), 
and of data provided by the NASA GALEX mission.
Part of the data were obtained through AAVSONet, run by the American Association of Variable Star Observers, \IBVSshortlink{http://www.aavso.org}{AAVSO}
through the Tzec Maun Foundation \IBVSshortlink{http://www.tzecmaun.org}{Tzec Maun}
and by using the Bradford Robotic Telescope. \IBVSshortlink{http://www.telescope.org}{BRT}

\references

Butters O.W., West R.G., Anderson D.R., et al., 2010, {\it A\&A} {\bf 520}, L10

Drake A.J., Djorgovski S.G., Mahabal A., Beshore E., Larson S., Graham M.J., Williams R., Christensen E., Catelan M., Boattini A., Gibbs A., Hill R., Kowalski R., 2009, {\it ApJ} {\bf 696}, 870

G\"ansicke B.T., Dillon M., Southworth J., et al., 2009, {\it MNRAS} {\bf 397}, 2170

Haro G., Luyten W.J., 1962, {\it Bol. Inst. Tonantzintla} {\bf 3}, 37

Jones D.H., Saunders W., Colless M. et al., 2004, {\it MNRAS} {\bf 355}, 747

Jones D.H., Read M.A., Saunders W. et al., 2009, {\it MNRAS} {\bf 399}, 683

Martin D.C., Fanson J., Schiminovich D., Morrissey P., Friedman P.G., Barlow T.A., Conrow T., Grange R., 
Jelinsky P.N., Milliard B., Siegmund O.H.W., Bianchi L., Byun Y.-I., Donas J., Forster K., Heckman T.M., 
Lee Y.-W., Madore B.F., Malina R.F., Neff S.G., Rich R.M., Small T., Surber F., Szalay A.S., Welsh B., 
Wyder T.K., 2005, {\it ApJ Letters} {\bf 619}, 1

Ochsenbein F., Bauer P., Marcout J., 2000, {\it A\&A Suppl.} {\bf 143}, 221

Savoury C.D.J., Littlefair S.P., Dhillon V.S. et al., 2011, arXiv:1103.2713v1 [astro-ph.SR]

Southworth J., Copperwheat C.M., 2011, arXiv:1101.2534v1 [astro-ph.SR]

Wils P., 2009, {\it IBVS} {\bf 5916}

Wo\'zniak P.R., Vestrand W.T., Akerlof C.W., et al., 2004, {\it AJ} {\bf 127}, 2436

\endreferences

\end{document}